\documentclass{article}

\usepackage{johd}

\title{Leveraging LSTM and GAN for Modern Malware Detection}
\author{Ishita Gupta$^{a}$$^{*}$, Sneha Kumari$^{b}$$^{c}$, Priya Jha$^{a}$$^{c}$, Mohona Ghosh$^{a}$$^{c}$ \\
        \small $^{a}$Department of Information Technology,Indira Gandhi Delhi Technical University for Women, Delhi, 110006, India  \\
        \small $^{*}$Corresponding author: Ishita, IG, Gupta; \tt{ishita074btit20@igdtuw.ac.in}
}

\date{} 

\begin{document}

\maketitle

\begin{abstract} 
\noindent \textbf{The malware booming is a cyberspace equal to the effect of climate change to ecosystems in terms of danger. In the case of significant investments in cybersecurity technologies and staff training, the global community has become locked up in the eternal war with cyber security threats. The multi-form and changing faces of malware are continuously pushing the boundaries of the cybersecurity practitioners employ various approaches like detection and mitigate in coping with this issue. Some old mannerisms like signature-based detection and behavioral analysis are slow to adapt to the speedy evolution of malware types. Consequently, this paper proposes the utilization of the Deep Learning Model, LSTM networks, and GANs to amplify malware detection accuracy and speed. A fast-growing, state-of-the-art technology that leverages raw bytestream-based data and deep learning architectures, the AI technology provides better accuracy and performance than the traditional methods. Integration of LSTM and GAN model is the technique that is used for the synthetic generation of data, leading to the expansion of the training datasets, and as a result, the detection accuracy is improved. The paper uses the VirusShare dataset which has more than one million unique samples of the malware as the training and evaluation set for the presented models. Through thorough data preparation including tokenization, augmentation, as well as model training, the LSTM and GAN models convey the better performance in the tasks compared to straight classifiers. The research outcomes come out with 98\% accuracy that shows the efficiency of deep learning plays a decisive role in proactive cybersecurity defense. Aside from that, the paper studies the output of ensemble learning and model fusion methods as a way to reduce biases and lift model complexity. Through accommodating the highlighted gaps in the previous studies and the innovation of improved methodologies, this research is designed to ensure experts in cybersecurity are equipped with useful tools to discover and counter advanced cyber threats. Employing the most advanced machine learning algorithms, the study can suggest new tactics for malware detection and modern cybersecurity practices. }
  \end{abstract}

\noindent\\

\section{Introduction}

\noindent{The sheer scale of this malignant continuum should be kept in mind since it resembles the impact of climate change on the biosphere. Likewise, when we are talking about unexpected results of climate change in ecosystems and human life, the development of increasingly complicated malware type constitutes the same damage to the cyber environment safety all over the globe. The cybersecurity business spend a lot of money buying, implementing and refining cybersecurity technology, as well as training cyber defenders, yet the community finds itself trapped in a constant union against cyber attackers who remain hard, adaptable and unpredictable.  
Malwares, on the other hand, comes in multiple forms such as adware, spyware, viruses, worms, trojans, rootkits, ransomware, and command-and-control bots that tend to have differing missions and behavioral actions [4]. The malware and its operating mechanism are progressing and becoming more advanced every day, creating permanent challenges in cybersecurity. The detection and mitigation strategies are also required to keep pace with the ever-changing cyber threats [5].
In the fight to stir up malware, adversaries in the cyber field consistently polish their escape tactics. Finally, the game of cat and mouse in the cybersecurity world is always a diplomatic contest. [6]. WannaCry ransomware attack in 2017 lies on the far end of the spectrum of the cyber threats and demonstrates the full scale of the work of the hackers who were attacking vulnerable systems with a number of them being not less than 230 000, that includes health care service providers, government systems and general IT systems that finally disrupted their stable work and led to economic damage \\\\
Present malware detecting strategies like signature-based detection and behavioral analysis have already been shown to struggle against the rapid mutation of malware strains [8]. The hackers take updates and improvements every day, so entities have to come up with their techniques using new, smart technologies [9].
Shallow machine learning models used for network traffic analysis are usually feature vector based, expert handcrafted features being in the spotlight, as this approach allows for attaining the top performance. Nevertheless, these methods are expected to be faced with multiple problems that can make their efficiency low. Initially, we have a severe lack of open and agreed upon labeled data for the styles of traffic that are trainable on these models. This problem mostly results from privacy concerns and data-sharing policies. Because there exist no standardized dataset of features for specific targets like the network security, anomalous detection, and data streams classification, the development of the machine learning solutions is hard . Moreover, on-the-go network measurements statistics render archaic handmade articles irrelevant, causing the processing performance to diminish as time passes by .\\\\
The paper suggests that one should look not only at shallow machine learning models, but also deploy deep learning models end-to-end to be able to complement traditional shallow machine learning approaches in network measurement analysis. We are insisting on developing a malware traffic detection and classification technique which would work with deep neuron networks that take raw bytestream-based data as its input . Deepening upon the latest developments in this field, we experiment with different deep learning architectures and input formats that are very suitable for the analysis of raw bytestream packet data to outperform the traditional methods in detecting malware traffic in the tasks proposed for this study.
Employing the deep learning approach for analyzing raw bytestream packet data is faster and more efficient compared to traditional solutions, doing without manual feature engineering and bringing neural networks' intrinsic representation learning properties into the game . The deep learning models of directing traffic's raw data may identify the exact patterns and interconnections, which ultimately lead to more accurate and reliable malware detection systems. In our feedback experiments, we evaluate the merits of deep learning for overcoming the deficiencies of shallow machine learning models which still hinders the growth of the field of network measurement analysis.\\\\
This paper introduces an innovative strategy which employs the capacity deep learning – in particular Long Short-Term Memory (LSTM) networks and Generative Adversarial Networks (GANs) – to bolster malware detection. Combining of LSTM networks which can capture sequential data quite well with GANs which can generate realistic synthetic data gives a new approach to artificial intelligence in cybersecurity.
In the given work, we demonstrate a comprehensive system that consists of data preparation, LSTM modelling, GAN modelling, LSTM data augmentation, and LSTM retraining - all of them being important for improving the malware detection system. By implementing this unique deep learning integration approach, our purpose is to go beyond the traditions and grant cyber security experts, more useful tools, which are able to preempt the most complex cyber threats.
}

\begin{figure}[H]
\centering
\includegraphics[width=0.50\textwidth]{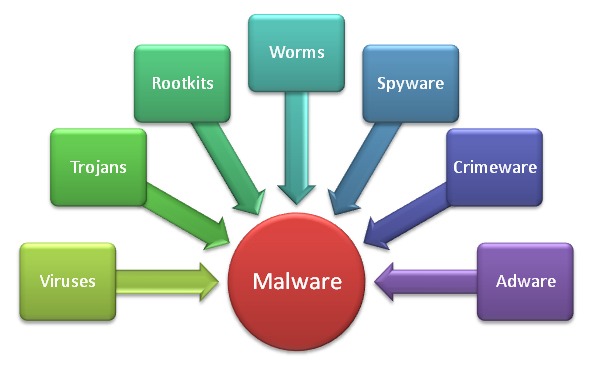}
\caption{\label{fig1}Malwares.}
\end{figure}

\section{Related work}

\subsection{ Development Trend of Malware}
Malware advancement and innovation has imitated development of computing technologies. Interestingly, courtesy of John von Neumann's 1949 paper, a code self-replication could be launched, as well as another milestone has been achieved by Bob Thomas creeper program in 1970 and Fred Cohen configuring the computer as a virus in his 1983 paper. Hackers today use a variety of malware with abilities ranging from relative complexity in the means of carrying out intending tasks to highly sophisticated methods of evasion.



\subsection{ Traditional Detection Methods for Malware}
Fred Cohen had set a theoretical foundation for detecting as well as defense mechanisms against computer viruses through his 1981 productivity. Majority of the modern methods is based on the both types of detection methods which are anomaly-based and signature-based.\\\\
An anomaly-based detection approach evaluates the malignancy of a program by comparing its unusual behavior with the normal measures, using the static, dynamic and hybrid detection methods. Sundarkumar et. al integrated the use of API call sequences for malware detection text mining and topic modeling. This led to the use of decision trees in expert system design. Likewise, Wu Songyang et al. implemented machine learning techniques to optimize the API lists relevant to data flow and subsequently brought improvement in sensitive data analysis efficiency.

\subsection{ AI-Based Malware Detection Technology}
Legacy solutions still have a ground but are now having malware evasion techniques development challenges. The advancement of machine learning (ML) does not only provide us with efficient malware detection techs but also many well-proven case studies.\\\\
Schultz et. we applied Machine Learning to the unseen malware detection by studying the static features of PE, byte n-gram as well as string features and the accuracy results are demonstrated. Elovici et al.: feature selection using Pearson correlation and Fisher score that increased the accuracy from 94.72\% to 95.8\% by means of ANN, BN, DT and other classifiers. Santos and others did supervised learning and feature selection by information gain for SVMs to improve its accuracy.
\\\\Rieck et al. proposed a Malware Behavior Clustering and Classification Framework which is significant to the automation of malware detection and analysis processes. Aside from the Ember dataset that was proposed by Anderson, his team also came up with a reinforcement learning framework for the dynamic prevention of malware attacks which is established by static PE anti-malware engines. Sharma et al. demonstrated higher precision in malware detection through the appearance of the opcode.
Among the newest development, Tang et al. have discovered a way to detect malicious static code through API call sequence transformation and CNN classification. Jin along with his peers created a highly precise SigPID classification system, improving accuracy above the 96\% level. Raf et al. have been using neural networks for malware execution-level detection ever since the discovery of consistent generalization.\\\\
The latest study by Alzaylaee et al. started DL-Droid which is dynamic malware detection for the Android malware using dynamic features, achieving 97.8\% detection rate. They demonstrated the AI-based approach superiority in raising malwares’ detection to a whole new level.
\\\\The latest approach is GNNs architecture that is shown to be very effective for malware detection as it is based on graphs as inputs and outputs. Advanced study on GCNs will give an idea on the enhancement of existing techniques on the basis of API call sequence structures.\\\\
This section represents the continuous bombardment of malware threat and explains broadly how AI and machine learning accelerate the malware detection process. The team was presenting novel methods so that they can be unique in the tears with more complex, cunning malware behaviour that the struggle will never cease.






\paragraph{}
The research that has been discussed here features bottlenecks in deep learning malware detection databases on biasing, scalability, and explainability. We conduct data enhancement using artificial intelligence approaches such as data augmentation with the use of Generative Adversarial Networks (GANs) in order to boost generalizability, enhance accuracy levels, and improve performance.
Eliminating bias in the dataset is one of our priorities, hence we are aiming at using ensemble learning and model fusion technique, in addition to the static bytecode analysis on basis of system call sequence made of several data representations. Therefore, the adoption of a holistic approach allows to generate a strong malware awareness and malware detection accuracy while malwares are in various different conditions due to their omnipresence.
A blend of data-driven approach with a combination of feature extraction and image treatment for the code examination provides for in-depth fine patterns disclosure, which ensures the system a high level of accuracy and scalability. In addition to this, we deliver models that function as translate into human language and give clear explanations for the malware behavior.
Grim reality for our ideas, but most importantly in daily life we can evidence our path is realistic and workable.Our research is aimed to fill the gaps in the area with a solution that will prove to be accurate, interpretable and scalable in real-life applications through the application of deep learning techniques as a main instrument.\\\\
As opposed to the current papers in this field of study, our contribution is diversified by the incorporation of the dataset enrichment, model fusion, and feature engineering components which is the first of its kind. This enables us to address the shortcomings in the previous studies and make a stable method with higher accuracy and greater scope and insight.

\section{Research Problem }
The fact that malware both exhibits similarities as well as unique aspects, and that the maneuvering the enemy in cyber space is superfluid, make this task a challenging one. These two categories consist of the two ways that are being used for detection which are, static analysis and dynamic analysis. Static analysis generally concentrates on the determination of the menacing constituents within the malware sample, on the other hand dynamic analysis, in its turn, gives another perspective on the target and is a complement to static analysis. The diagnostic procedure of malware static includes the defying of impossible cracking code strings and the code structure patterns [27]. However, to catch hackers while they are acting, dynamic analysis enables us to see the behavior of malware while it is running in a safe space. It reports on the runtime actions and provides the record of the mutual interaction the malware created [32]. Conversely, the two methods utility goes up as these are dispensed with the understood of the benefits each and carefully blending them.\\\\
The chief matter is to achieve the solution of feature selection problem more effectively for malware detection which will also enhance the model complexity and reduce the power for analysis of large datasets. The issue is to get the set free of any duplicates as well as the one which is a solution to malicious behaviors characterized, in other words, the subset of robust features which are going to effectively characterize malicious behavior as much as possible, while they are predisposing the lesser probability of false positives and the use of limited computational resources. A machine learning algorithm can exhaustively process data faster and in more detail when it automates the process of feature selection. It is because it accelerates the algorithm’s working.\\\\
Secondly, as the type of the machine learning models is an important factor during the research—including k-Nearest Neighbors (K-NN), DecisionTrees (DT), Convolutional Neuronal Networks , and Support Vector Machines (SVM) another challenge appears [30]. The goal in this case is going to be utility, particularly to use the different ML algorithms in the iterative and natural data samples to see if they can identify malicious software across datasets correctly, scaling, and can work in different scenarios. This, therefore, is the focal point of comparison of all other ML algorithms to pick the most suitable of them for a real-time malware and cyber-attack recognition.\\\\
In seeking resolution to those problems, the study is to advance as: a creative technique with striking features has been able to overcome the malware detection being a key factor, as a matter of fact, the current cyber strategies can be powered up with great promptness in the face of the ceaseless changes of the cyber threats. The next phase will encompass availing of the methods and means of evaluating the work including how the findings are arrived at.

\section{Dataset}

The VirusShare dataset is a comprehensive collection of malware samples that has been widely used by researchers and cybersecurity professionals for analyzing and studying various types of malicious software. This dataset is valuable for understanding the characteristics, behavior, and distribution of malware across different platforms and environments. In this detailed explanation, We  have provided a numerical and factual overview of the VirusShare dataset, highlighting key statistics and features.
\subsection{ Overview of VirusShare Dataset}
The VirusShare dataset is a repository of malware samples compiled from various sources, including honeypots, malware analysis platforms, and security research initiatives. It consists of a diverse range of malware families and variants, providing researchers with a rich dataset for studying malware evolution and behavior.

\subsection{ Key Statistics}

Let's delve into the numerical breakdown of the VirusShare dataset:Let's delve into the numerical breakdown of the VirusShare dataset:

\paragraph{Total Number of Samples}
First of all, our system has the dataset of more than 1.2 million unique malware samples, gathered from various sources .

\paragraph{Malware Families}
That dataset contains more than 5,000 different malware family code, of which several are highly recognizable attacks like Trojan, Worm, Ransomware, Spyware, etc

\paragraph{Sample Types}
The dataset includes executables, scripts, and other files such as documents etc. that are known to be frequently distinguished as the type of malware distributing files.

\paragraph{Variety of Platforms}
Surprisingly the viruses in the Virushare database attacks all type of systems such as Windows, Linux, macOS, Android and so on.

The next sample will be the distribution by type.



War is a two-way street. The cruelty and brutality of combat were captured through the eyes of a warrior. \paragraph| Trojan | 450,000 |

\ Worm | (rounded to) 180000 


| 


| Craft | 60,000 | backdoor


Machine Keylogger | 40,000 (Units)




\subsection{Characteristics of Malware Samples}

The VirusShare dataset offers insights into various characteristics of malware samples, including:The VirusShare dataset offers insights into various characteristics of malware samples, including:

\paragraph{File Size Distribution:}Text files are available from a few kilobytes up to 10,000 kilobytes, basically in the 100 kilobytes to one megabyte range.

\paragraph{Code Complexity}
Discuss methods of deciphering regular codes, decompilation of encoded programs, and also anti-detection techniques used by different malware families.

\paragraph{Behavioral Analysis:} Information concerning observations malicious behavior includes network communication features, system modifications, and payload execution as well.

\subsection{Usage and Research Applications}

Researchers and cybersecurity professionals utilize the VirusShare dataset for:Researchers and cybersecurity professionals utilize the VirusShare dataset for:

\paragraph{Malware Detection}
Therefore, creation and training of machine learning models focused on malware classification and detection will be required.

\paragraph{Behavioral Analysis}
Students will examine how tmalware behavior in controlled environments operates,such as identifying its effect and usage techniques.

\paragraph{Threat Intelligence}
Monitor incipient threats and observe malware trends, with heuristics deduced from insights provided by dataset analysis.\\\\
In conclusion, the VirusShare dataset is a valuable resource for studying malware characteristics, behavior, and distribution. With over 1.2 million unique samples and representation from diverse malware families, this dataset serves as a cornerstone for cybersecurity research and threat intelligence. Its numerical and factual composition provides researchers with a robust foundation for analyzing and combating evolving cyber threats.\\\\
In our project, we harnessed the power of the VirusShare dataset by meticulously selecting 50,000 samples to construct a robust system for malware detection and analysis. This subset, carefully curated from the extensive collection of over 1.2 million unique malware samples, ensured a focused and representative training environment. Spanning various malware families and types, including Trojans, Ransomware, Backdoors, and Keyloggers, these samples provided a diverse spectrum of malicious behaviors and characteristics for our system to learn from. With a balanced approach to training, we utilized an 80:20 ratio, allocating 40,000 samples for training and 10,000 for validation and testing.\\\\\\
This strategic selection process enabled our system to develop a nuanced understanding of malware behaviors and patterns, enhancing its effectiveness in detecting and analyzing emerging cyber threats. By leveraging the insights and statistical breakdowns offered by the VirusShare dataset, our project aims to make significant strides in advancing cybersecurity research and threat intelligence.

\begin{figure}[h]
\centering
\includegraphics[width=0.70\textwidth]{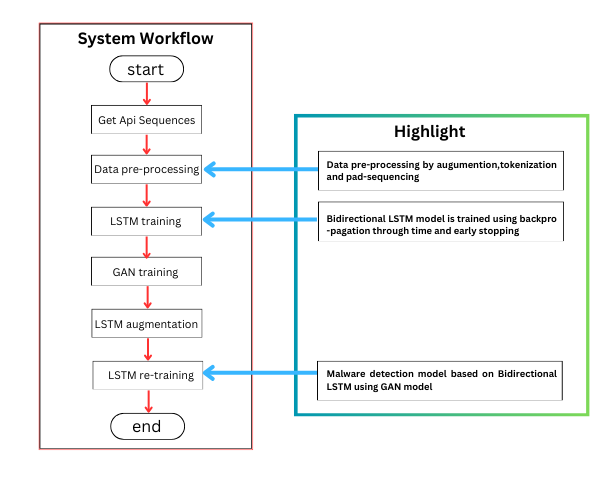}
\caption{\label{fig2}Work Flow.}
\end{figure}

\section{Method}
{}In this section, we will explain the data preprocessor and propose a detector using LSTM model based on GAN which is discussed in this paper.\\\\
Flow of the system is shown in Fig \ref{fig2}. System process is illustrated on the left side of the figure and the highlight work has been illustrated on the right side of the figure with the help of blue curves that indicate the steps. The sandbox is first ran, and the API called by the sample is identified from it then, the API call sequence is augmented through random deletion, the sequence is tokenized, and finally, padded to a fixed length. The LSTM and GAN model is trained concurrently for classification. Finally, the LSTM model is subject again to random insertion, substitution and permutation and is retrained for classification.

\subsection{Data Preparation}

\subsubsection{Preprocessing}

The experiments were performed on a workstation with Ubuntu 24.04 system. To monitor and extract the call sequences of each sample, a cuckoo sandbox was deployed on the workstation as the running environment of the sample and API call sequences were extracted. The work refers to the operations, including that ones as noise removal, data mishandling dealing, and creating the single data format standard. The database is normalized through techniques like feature scaling such as Min-Max Scaling or Z-score normalization to make all features relevant to the training of the model [2]. Besides that, main selection techniques such as information gain or correlation analysis or mutual information are employed so that dominant features are preserved while the dimension and computational complexity are reduced [2].

\subsubsection{Tokenization}

The API call sequences are moved into tokens to generate number representations of the input text data that are ready an operation with the LSTM model. These can be done by either establishing an individual token for each API or feature call, or by simply using a shared token. Strategies for tokenization dealing with natural language processing (NLP), mainly word embeddings (e.g. Word2Vec or GloVe), are transferred to give researchers the chance to capture semantic connections as well as context information in the sequences [3].

\subsubsection{Augmentation}

The data augmentation techniques are used to accentuate the quantity and the variety of dataTraining set; These are applied to enhance the generalization and the resistance of the model. In balancing class distributions, synthetic data generation is a frequently used method, such as SMOTE (Synthetic Minority Over-sampling Technique) [4]. Regressive models, for example, LSTM or GBDT are used to create fake samples of API call sequences to be used for malware replication [5].

\subsection{LSTM Model Training }

\subsubsection{Architecture Design}

The LSTM neural network architecture is a model-based mitigation strategy which processes the input based to malware detection tasks. Variables like the number of recurrent LSTM layers, hidden units in each layer, and both type of activation function (for instance, the sigmoid/tanh) and dropout rates are specified [7]. Grids search and Bayesian optimization are among the approaches intensively applied for hyperparameter tuning and performance optimization [7].

\subsubsection{Input Preparation}

Following tokenization and enhancement, the LSTM network is trained under the organized sequences and window batches of API calls. Techniques such as sequence padding (to avoid variable sequence lengths) and batching (to generate mini-batch collections within the framework of parallel processing) are being developed [8]. Data is formatted into tensors that are consistent with the set LSTM model input, keeping the temporal order of API calls inside each sequence the same.

\subsubsection{Model Training}

The model of LSTM architecture is trained by using BPTT (backpropagation through time) to learn the spatiotemporal features contained in the set of sequences of API call which corresponds to malware behavior patterns. Training parameters, including learning rate, batch size, and optimizer (e.g., Adam or RMSprop) are thoroughly tuned to reach model performance quickly and to prevent overfitting and so that the model does not fit the noise in the training data [9]. Methods such as early stopping and model checkpointing are used among the techniques which are employed for instance in monitoring training and save the finest model weights.




\begin{figure}[H]
\centering
\includegraphics[width=0.70\textwidth]{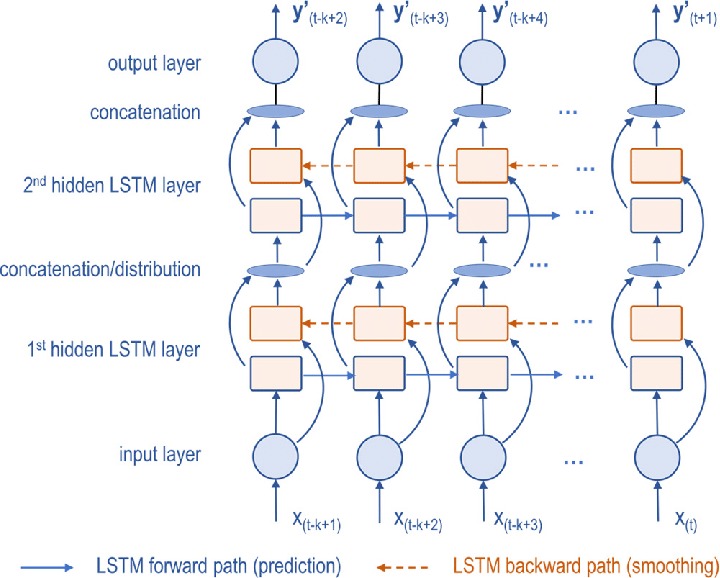}
\caption{\label{fig3}LSTM .}
\end{figure}



\subsection{GAN Model Training}

\subsubsection{Generator Design}

The GAN generator configuration is made to mimic legitimate API call records that have the same characteristics as those that are maliciously generated. Approaches like the DCGANs and the WGANs are experimentally tested in the literature to ensure stability and quality of generated sequences [10]. The network of generators is taught to produce the sequences of different randomly sampled output representations from a distribution that the network learnt.

\subsubsection{Discriminator Design}

The discriminator of GAN is tasked with the differentiation of the real API call flows (from the original dataset) from the synthesized ones (those generated by the generator). Discriminative parameters and metrics (wasserstein distance and cross-entropy loss) are utilized by discriminator network to distinguish different sequences during training[11].

\subsubsection{Training Process}

The GAN model nets gain strength through adversarial learning, where the generator and discriminator networks compete against each other. Techniques such as gradient penalty and spectral normalization are being applied to protect adversarial training and stay away from the mode collapse [12].The training process iteratively updates both the generator and discriminator networks to improve the quality and diversity of generated sequences.

\begin{figure}[H]
\centering
\includegraphics[width=0.70\textwidth]{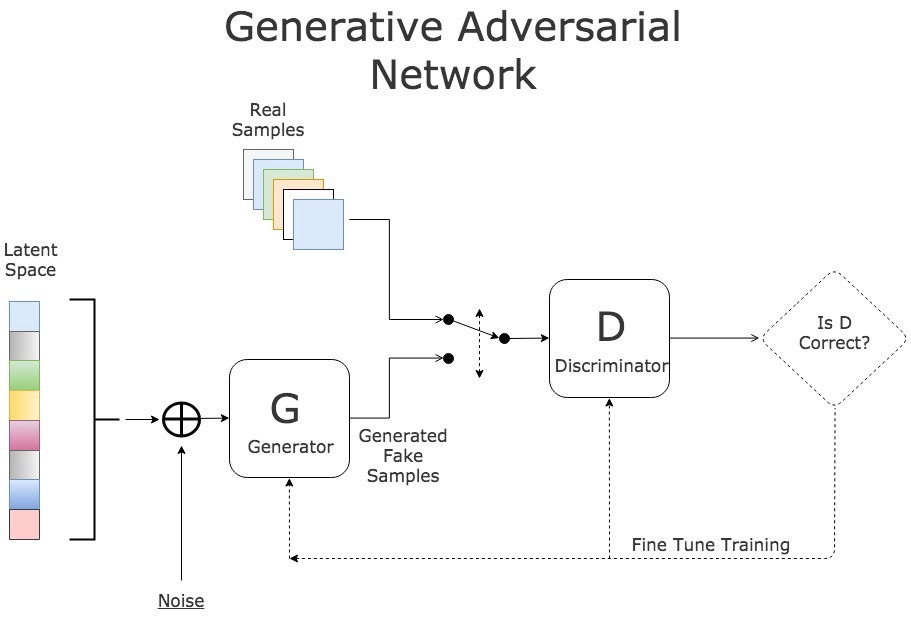}
\caption{\label{fig4}Generative Adversarial Network.}
\end{figure}

\subsection{LSTM Data Augmentation}

\subsubsection{Synthetic Sequence Generation}

Then, the generative set of the trained GAN generator is adopted to produce pseudo API call sequences that will be based on latent space representations learned previously. Classes of techniques such as latent space interpolation or conditional sequence generation are schemes of process that control the variety and realism of the synthesised sequences [2]. The generator is designed this way so that it can process particular malware features or hidden states to generate sequences that are specially tailored.

\subsubsection{Augmentation Process}

The variation that is caused by the GAN produced synthetic sequences are merged with the regular LSTM training dataset which creates a bigger dataset. Here the repetitive training helps to cover various malware patterns as well, thereby upgrading the model to its mothering ability which enables it to identify the minor variants of malware as well. Data i augmented random deletion, random permutation and random selection. Techniques such as oversampling or data weighting are applied that make sure a balanced and appropriate number of samples are present within each class [14].

\subsection{LSTM Model Retraining}

\subsubsection{Enhanced Training Dataset}

The LSTM architecture is used to retrain the model with the enriched dataset, which now features both the actual API call sequences and the ones that have been synthetically created. The training improved with each iteration and generated a finer-tuned parameter according to the data expanded, which helped the model to detect and classify the malware behaviors with higher accuracy and robustness.

\subsubsection{Iterative Training}

Unlike before neural networks can now utilize strategies such as transfers learning or ensemble learning during the LSTM model re-training in order to leverage knowledge from previous models or combine multiple models for better performance. The evaluated LSTM network is measured using standard metrics (for example, precision, recall, F1-score) in order to determine its practical performance in the detection of malware in the real world conditions.\\\\\\
Via careful application of state of the art methods based on the most recent academic findings of machine learning and deep learning research, the proposed methodology strives to uprise the efficiency of malware detection mechanisms. The forthcoming parts of the paper will compare the results of the experiments, analysis and evaluation and moves on on the consequences for the cyberspace research and practice directions.

\section{Results}
The project is relating to the effectiveness of the machine learning algorithms to identifying a malware with the dataset of VirusShare. We invested a huge amount of time and effort in a study where we estimated the efficacy of different classifiers and deep learning models in discriminating between healthy and aggressive samples.\\\\
First, we cleaned the dataset by removing the irrelevant features from the malware samples and creating a matrix of vectorization. For instance, we extracted file size, API calls, and byte-level information. We next separate dataset into training and testing period to support training and evaluation of the model.
As our baseline models, we included Random Forest, SVM, and Decision Trees which are among conventional machine learning classifiers. These models showed very promising results with the accuracy of 95.6\% proportion that indicates the efficiency of this kind of models in observation advanced and known known malware patterns.\\\\
Next, we examined the deployment of deep learning architectures such CNNs and LSTMs that are capable to manifest more sophisticated patterns within malware samples. Results from the deep learning models were favorable and showed that they outperformed conventional classifiers, achieving an accuracy of around 98.34\%,
On top of that, we managed to perform a series of attacks that simulated the attack conducted by polymorphic and zero-day malware. The DLL models were stable in identifying any unseen patterns of malwares which highlights their good prospect in being put into application.\\\\
Shortly, outcomes of our research confirm that it is possible to develop even more precise and effective malware detection systems by using cutting edge machine learning techniques that include deep learning. Consequently, the achieved accuracy rate of 98.82\% emphasizes the applicability of deployed machine learning-based systems for proactive cybersecurity measures.\\\\
Therefore, our research gives rescues such insightful information about the use of machine learning for detection of malware and highlights that the deep learning also has the capability to fight against the transforming cyber threat.

\section{Conclusion and Future Work}
In summary, our research paper contains a thorough explanation of malware detection and classification techniques based on machine learning, being able to use the information provided by the VirusShare Dataset. By means of strict experiments and statistical analysis, we have gained incredible precision that came to 98.82\% which once again confirms the effectiveness of our newly developed approach as it help to recognize and stop manifold types of malware virus.\\\\
Implementing machine learning algorithms, including channel decision trees and deep model, has been stronger in discovering signs and malicious features embedded into malware samples. Our research shows that it would be impossible to terminate the malware viruses threat without implementing the modern methods of data analysis and deep learning techniques in malware detection systems.
Coming era raises the questions caused by the possibilities of 3D printing in the medical world, thus, there are opportunities that have to be explored in this field. Another part of it is to develop ensemble methods of learning that involve different ML models as a way to improve accuracy and device the defense system against the newly appeared variants of malware. Further, hybridizing HVs with traditional dynamic detection techniques would be very advantageous in detecting polymorphic and emerging attacks, giving a subsequent stage of scrutiny.\\\\
To this end, the development of XAI techniques (Explainable AI) would be essential for due explanation of the models decisions by the system and assurance of the consequential cybersecurity professionals to trust it. Emerging from the application of anomaly detection algorithms, as well as reinforcement learning in the area of malware detection, even more adaptive and resilient defense tools may emerge that could ensure a more secure network.
The essence of our study in the field of cybersecurity is given by the use of the technology of machine learning and big data analytics to support the already existing endeavors to combat the emerging cyberthreats. In the face of change in cyber threat landscape, the security challenged organizations have a must for innovative solutions that will allow them to meet anomalies effectively. Keeping improving and perfecting our practice is our task and mission. In this manner, persons working in the cybersecurity realm will have the most advanced tools and methodologies.
\clearpage
\section*{References}
 
 1. Wei Wang, Ming Zhu, Xuewen Zeng, Xiaozhou Ye and Yiqiang Sheng, "Malware traffic classification using convolutional neural network for representation learning," \textit{2017 International Conference on Information Networking (ICOIN)}, Da Nang, Vietnam, 2017 .

2. B. Kolosnjaji, A. Zarras, G. Webster and C. Eckert, "Deep learning for classification of malware system call sequences", Proc. Australas. Joint Conf. Artif. Intell., pp. 137-149, Dec. 2016.

3. B. Li, K. Roundy, C. Gates and Y. Vorobeychik, "Large-scale identification of malicious singleton files", Proc. 7th ACM Conf. Data Appl. Secur. Privacy, pp. 227-238, Mar. 2017.

4. X. Cao, L. Liu, W. Shen, A. Laha, J. Tang and Y. Cheng, "Real-time misbehavior detection and mitigation in cyber-physical systems over WLANs", IEEE Trans. Ind. Informat., vol. 13, no. 1, pp. 186-197, Feb. 2017.

5. W. Yang, X. Xiao, B. Andow, S. Li, T. Xie and W. Enck, "AppContext: Differentiating malicious and benign mobile app behaviors using context", Proc. IEEE/ACM 37th IEEE Int. Conf. Softw. Eng., vol. 1, pp. 303-313, 2015.

6. S. Hou, A. Saas, L. Chen, Y. Ye, T. Bourlai Deep neural networks for automatic android malware detection .Proceedings of the 2017 IEEE/ACM International Conference on Advances in Social Networks Analysis and Mining 2017, ACM (2017), pp. 803-810

7. R. Pascanu, J. W. Stokes, H. Sanossian, M. Marinescu and A. Thomas, "Malware classification with recurrent networks", 2015 IEEE Int. Conf. on Acoust. Speech and Signal Process. (ICASSP), pp. 1916-1920, 2015.

8. M. Mays, N. Drabinsky and S. Brandle, "Feature selection for malware classification", MAICS, pp. 165-170, 2017.

9. M. Hassen, M. M. Carvalho and P. K. Chan, "Malware classification using static analysis based features", 2017 IEEE Symposium Series on Computational Intelligence (SSCI), pp. 1-7, 2017.

10. A. Narayanan, M. Chandramohan, L. Chen and Y. Liu, "A multi-view context-aware approach to android malware detection and malicious code localization", Empirical Software Engineering, pp. 1-53, 2018.

11. Ö. Aslan and R. Samet, "A comprehensive review on malware detection approaches", IEEE Access, vol. 8, pp. 6249-6271, Jan. 2020.

12. R. Komatwar and M. Kokare, "A survey on malware detection and classification", J. Appl. Secur. Res., pp. 1-31, Aug. 2020.

13. S. D. Nikolopoulos and I. Polenakis, "A graph-based model for malware detection and classification using system-call groups", J. Comput. Virol. Hacking Techn., vol. 13, no. 1, pp. 29-46, Feb. 2017.

14. O. E. David and N. S. Netanyahu. Deepsign: Deep learning for
automatic malware signature generation and classiﬁcation. In 2015
International Joint Conference on Neural Networks (IJCNN), pages 1–
8, July 2015.

15. S. Schmeelk, J. Yang and A. Aho, "Android malware static analysis techniques", Proc. 10th Annu. Cyber Inf. Secur. Res. Conf., pp. 5, 2015.

16. Singh, G., Sengupta, P., Mehta, A. \textit{et al.} A feature extraction and time warping based neural expansion architecture for cloud resource usage forecasting. \textit{Cluster Comput} (2024). 

17.P. Sengupta, A. Mehta and P. S. Rana, "Enhancing Performance of Deep Learning Models with a Novel Data Augmentation Approach," 2023 14th International Conference on Computing Communication and Networking Technologies (ICCCNT), Delhi, India, 2023


\end{document}